\documentclass[twocolumn,showpacs,preprintnumbers,amsmath,amssymb,epsfig]{revtex4}

\usepackage{graphicx}% Include figure files
\usepackage{dcolumn}% Align table columns on decimal point
\usepackage{bm}% bold math
\usepackage{epsfig}

\def\fun#1#2{\lower3.6pt\vbox{\baselineskip0pt\lineskip.9pt
  \ialign{$\mathsurround=0pt#1\hfil##\hfil$\crcr#2\crcr\sim\crcr}}}

\input epsf

\def\be{\begin{equation}}
\def\ee{\end{equation}}
\def\ba{\begin{eqnarray}}
\def\ea{\end{eqnarray}}
\def\half{\frac{1}{2}}
\def\nn{\nonumber}

\newcommand{\dbd}[2]{\ensuremath{\frac{\ds #1}{\ds #2}}}
\newcommand{\ds}{\mathrm{d}}
\begin{document}

\preprint{}

\title{Large Scale Structure Formation of Normal Branch in DGP Brane World Model}

\author{Yong-Seon Song}
\email{yong-seon.song@port.ac.uk}
\affiliation{Institute of Cosmological $\&$ Gravitation, 
University of Portsmouth, Portsmouth, PO1 2EG, UK }

\date{\today}

\begin{abstract}
In this paper, we study the large scale structure formation of 
the normal branch in DGP model
(Dvail, Gabadadze and Porrati brane world model)
by applying the scaling method developed by Sawicki, Song and Hu
for solving the coupled perturbed equations of motion 
of on-brane and off-brane.
There is detectable departure of perturbed gravitational potential 
from LCDM even at the minimal deviation of 
the effective equation of state $w_{\rm eff}$ below $-1$.
The modified perturbed gravitational potential 
weakens the integrated Sachs-Wolfe effect which is strengthened
in the self-accelerating branch DGP model.
Additionally, we discuss the validity of the scaling solution
in the de Sitter limit at late times.
\end{abstract}

%\pacs{draft}

%\keywords{CMB-inflation}

\maketitle

\section{introduction}
Dvali, Gabadadze and Porrati proposed a brane world model embedded in
the Minkowski 5D dimension~\cite{dvali00},
which has been taken as
an alternative theory to explain cosmic acceleration.
There are two distinct branches of this model: 
one is the self-accelerating branch
(hereafter sDGP) and the other is the normal branch (hereafter nDGP).

Deffayet discovered the accelerated phase in sDGP 
without cosmological constant~\cite{deffayet00},
which has been tested geometrically
by many authors~\cite{dvali03a,Avelino01,fairbairn05,Maartens06,Song06}.
The detectability will be enhanced by the structure formation test
~\cite{song04,Schimd04,gabadadze04,ishak05,Knox:2006fh,linder05,gabadadze05,Song06sa,tang06,huterer06,Carroll06,Stabenau06,Pal06hg,Pal07,Hu07}.
The distinct evolution of perturbed gravitational potential of sDGP
has been studied by many authors at scales 
in quasi-static limit~\cite{deffayet02a,lue04,koyama05,Sawicki06}
and scales at nearly horizon size~\cite{Sawicki06}.
sDGP strengthens the integrated Sach-Wolfe (hereafter ISW) effect,
and leaves a detectable signature on CMB power spectra
and ISW-galaxy cross-correlation~\cite{Song06}.

There is no cosmic acceleration induced spontaneously in nDGP
without introducing cosmological constant.
But it's attracting features are that nDGP has a spectrum leading to the
stability in quantum level~\cite{Charmousis06}, 
and that the effective equation of state
$w_{\rm eff}$ crosses $w_{\rm eff}=-1$ without phantom dark energy.
nDGP has been tested by many authors geometrically
as a natural theory crossing $w_{\rm eff}=-1$ 
line below~\cite{Sahni02,Chimento06,Lazkoz06}.
But no detailed study has been done 
to show how to formulate the large scale 
structure formation of nDGP in order to improve the detectability.

We study the large scale structure formation of nDGP
by using the scaling method developed by Sawicki, Song and Hu~\cite{Sawicki06}.
On-brane equations of motion are not in the closed form
without the information of the gradient crossing the bulk direction.
Off-brane equation of motion should be coupled in order to supply this gradient
of perturbations at the location of brane.
We assume the scaling ansatz for the perturbation on the brane,
and solve it iteratively until it converges.
We find that the quasi-static limit is valid at scales relevant to
ISW-galaxy cross-correlations.
The modified perturbations weaken the ISW effect,
which is the opposite behavior of perturbations found in sDGP.
While we are preparing this paper, Cardoso, Koyama, Seahra and Silva
independently derives the identical solution by using different 
approach~\cite{Cardoso07}.
They verify that the scaling method is valid for both branches of DGP.

\section{linear perturbation equations of normal branch DGP}
Thin 4D brane is embedded in the 5D gravitational theory in the DGP
brane model. The gravitational interaction propagates through the extra
dimension, while all other physical interactions are confined on the brane.
Tension is added to the matter-radiation energy on the brane
in order to accelerate the cosmic expansion in nDGP at late times.
The Einstein action is written as,
\ba
S&=&-\frac{1}{2\kappa^2}\int d^5x \sqrt{-g}R^{(5)}
-\frac{1}{2\mu^2}\int d^4x \sqrt{-\tilde{g}}\tilde{R}^{(4)} \nn \\
&+&\int d^4x \sqrt{-\tilde{g}} ({\sc L}_m+{\sc L}_{\sigma})\,,
\ea
where ${\sc L}_m$ denotes Lagrangian for matter-radiation energy,
${\sc L}_{\sigma}$ denotes Lagrangian for brane tension,
and bulk remains empty.
The gravitational interaction scales are $\kappa^2$ for 5D
and $\mu^2$ for 4D. 
The 5D gravitational interaction scale $\kappa^2$ is 
a free parameter to be tuned
to generate the observed cosmic acceleration.
The ratio between both gravitational scales is defined by 
$r_c=\kappa^2/2\mu^2$, which determines the critical scale of transition
between 4D and 5D gravity.

The modified Friedman equation in DGP is given by
\ba
H^2-\varepsilon \frac{H}{r_c}=\frac{\mu^2}{3}(\rho_m+\rho_{\sigma})\,,
\ea
where $\rho_m$ denotes the matter density and $\rho_{\sigma}$ denotes
the brane tension.
There are two alternative choices for the sign convention of $\varepsilon$,
$\varepsilon=\pm 1$.
The choice of $+$ sign is called as a self-acceleration branch
which self-accelerates the cosmic expansion without cosmological constant.
The other choice of $-$ sign is called as a normal branch.
There is no self-acceleration in this branch without cosmological constant.
In order to fuel cosmic acceleration,
it should be aided by a term like the brane tension working as
cosmological constant.

Although it is less attractive to take nDGP as an alternative candidate
replaces LCDM,
there is an interesting feature in this theory.
The effective equation of the state of effective dark energy $w_{\rm eff}$
is less than $-1$. The formulation is given by~\cite{Lazkoz06},
\ba
w_{\rm eff}=-1-\frac{\sqrt{\Omega_{r_c}}\Omega_m/a^3}
{[\Omega_{\sigma}-2\sqrt{\Omega_{r_c}}E(a)][\sqrt{\Omega_{r_c}}+E(a)]}\,,
\ea
where $\Omega_{r_c}=1/4H_0^2r_c^2$, and the expansion history factor
$E(a)=H(a)/H_0$ is
\ba
E(a)=\sqrt{\Omega_m/a^3+\Omega_{\sigma}+\Omega_{r_c}}-\sqrt{\Omega_{r_c}}\,.
\ea
5D gravitational effect screens cosmological constant,
which leads nDGP crossing $w_{\rm eff}=-1$ line.

The current constraint on the lower bound of $w_{\rm eff}$
is as low as $w_{\rm eff}\sim-1.3$ at a 95$\%$ confidence level
by the combined test of Supernovae and WMAP~\cite{riess04,spergel03}.
We vary $w_{\rm eff}$ from $-1.03$ to $-1.12$.
The cosmological parameters we use are:
$(\omega_b=0.025, \omega_m=0.128, H_0=72, 10^{10}A_{s}=21.2, n_s=0.95, \tau=0.09)$ from WMAP best fit~\cite{spergel03}.
The energy density of the brane tension $\Omega_{\sigma}$
is varied with the demanding value of $w_{\rm eff}$.

We write the equations of motion of nDGP.
There are two sets of equations. One is on-brane equations which are
the projected Einstein equations and the conserved equations on the 4D brane.
The other is off-brane equation which is the propagation equation
of perturbations through the bulk.

\subsection{On-brane perturbed equations of motion}
The perturbed metric on 4D brane is given by the Newtonian gauge,
\ba
 d s^2  = -(1+2\Psi)d t^2 + a^2(1+2\Phi)d{\bf x}^2
 \, .
\ea
The projected Einstein equations on 4D brane 
and the conserved equations give the on-brane equations of motion.
The projected Einstein equations provide a couple of constraint equations;
the Poisson equation and the anisotropy stress equation.
There are two conservation equations of matter and Weyl fluid.
Those equations complete the on-brane equations of motion.

The Poisson equation is given by,
\ba 
\frac{k^2}{a^2}\Phi = \frac{\mu^2\rho_m}{2}
\frac{2Hr_c}{2Hr_c+1}\Delta_m
-\frac{\mu^2\rho_m}{2}\frac{1}{2Hr_c+1}\Delta_E \,,
\label{e:poisson}
\ea
where $\Delta_m$ and $\Delta_E$ denote the comoving perturbed energy density
of matter and Weyl fluid respectively
($\Delta_m=\delta_m-3Hq_m$ and $\Delta_E=\delta_E-3Hq_E$).
Signatures of terms with $r_c$ are reversed in nDGP.
The geometrical anisotropy stress is generated by Weyl fluid anisotropy
fluctuations,
\ba
\Phi+\Psi = 
 \frac{1}{2r_c H \left(1+\frac{\dot{H}}{2H^2}\right) +1}
\frac{\mu^2 \rho_m}{3}a^2 \pi_E \,. \label{e:aniso}
\ea
Again, the overall sign is flipped in nDGP anisotropy stress equation.

The conservation equations of matter fluctuations are given by,
\ba
\dot{\delta}_m-\frac{k^2}{a^2}q_m=-3\dot{\Phi}
\ea
\ba
\dot{q}_m=-\Psi\,.
\ea
The conservation equations of Weyl fluid are given by,
\ba
\dot{\delta}_E+H\delta_E-\frac{k^2}{a^2}q_E=0
\ea
\ba
\dot{q}_E+\frac{1}{3}\delta_E-\frac{2}{9}\frac{k^2}{a^2}\pi_E=S\, ,
\ea
where the source term is,
\ba
S\equiv-\frac{2r_c\dot{H}}{3H}
\left[\frac{\Delta_m+\Delta_E}{2Hr_c+1}+
\frac{k^2\pi_E/3}{1+2Hr_c\left(1+\frac{\dot{H}}{2H^2}\right)}
\right]\,.
\ea

A complete set of on-brane equations of motion 
is given by combining all of these equations.
Unfortunately, these on-brane equations of motion are not in
close form, since the anisotropy equation is unknown due to
the undetermined $\pi_E$ on-brane.
We will show how to derive $\pi_E$ from the off-brane equation of motion
which is called the master equation.

\subsection{Off-brane perturbed equation of motion}
In the maximal symmetry of five-dimensional metric,
scalar perturbations can be expressed by a single scalar field
called as master variable $\Omega$.
The propagation equation of $\Omega$ through the bulk is 
given by~\cite{mukohyama00}
\ba\label{eq:master}
-\left(\frac{1}{nb^3}\dot{\Omega}\right)^.
+\left(\frac{n}{b^3}\Omega'\right)'
-\frac{n}{b^5}k^2\Omega=0\,,
\ea
where the 5D metric is 
$d^2s=-n(y,t)^2dt^2+b(y,t)^2d{\bf x}^2+dy^2$.
The components of the metric of nDGP are given by
\ba
n(y,t)&=&1-\left(\frac{\dot{H}}{H}+H\right)\mid y \mid \nn\\
b(y,t)&=&a\left(1-H\mid y \mid \right)\,.
\ea

It is clear to see how the off-brane perturbed equation of motion
helps us in solving the on-brane equations of motion by expressing
Weyl fluid fluctuations in terms of $\Omega$,
\ba\label{eq:weyl}
\mu^2\rho_m\delta_E&=&-\frac{k^4}{3a^5}\Omega \nn \\
\mu^2\rho_mq_E&=&-\frac{k^2}{3a^3}\left(\dot{\Omega}-H\Omega\right) \\
\mu^2\rho_m\pi_E&=&-\frac{1}{2a^3}
\left(\ddot{\Omega}-3H\dot{\Omega}+\frac{k^2}{a^2}\Omega
+\frac{3\dot{H}}{H}\Omega'\right)\,,\nn 
\ea
where $'$ denotes the derivative in terms of $y$.
The missing information on the brane
is the gradient of $\Omega$ crossing the bulk
in $\pi_E$ equation.
This gradient is supplied by solving Eq.~\ref{eq:master},
the propagation equation of perturbations along the bulk.
The coefficient of $\Omega'$ term has the hidden $\varepsilon$ 
which should be reversed in nDGP.

The perturbed Weyl fluid conserved equation can be re-expressed 
by using Eq.~\ref{eq:weyl} in terms of $\Omega$,
\ba\label{eq:Bian}
\ddot{\Omega}&-&3HF(H)\dot{\Omega}\nn\\
&+&
\left(F(H)\frac{k^2}{a^2}-\frac{H}{K(H)r_c}-\frac{2Hr_c+1}{r_c}RH\right)\Omega
\nn\\
&=&\frac{2a^3}{k^2K(H)}\mu^2\rho_m\Delta_m\,,
\ea
where $R$ denotes the gradient crossing the bulk at the location of brane,
\ba
R\equiv\frac{1}{H\Omega}\frac{\partial\Omega}{\partial y}\mid_{y=0}\,,
\ea
and the coefficients $F(H)$ and $K(H)$ are
\ba
F(H)&\equiv&\frac{1+2Hr_c\left(1+\frac{\dot{H}}{3H^2}\right)}{2Hr_c+1}\nn\\
K(H)&\equiv&\frac{2Hr_c+1}{1+2Hr_c\left(1+\frac{\dot{H}}{2H^2}\right)}\,.
\ea
We complete our formulations for nDGP by switching all signatures 
in the terms including $\varepsilon$.

\section{Scaling solution}
The complete set of equations of motion can be solved by using
scaling method developed by Sawicki, Song and Hu~\cite{Sawicki06}.
The evolution of $\Omega$ is parametrized by the exponent
of $a$ based upon the assumption of scaling ansatz on the brane,
which enable us to intercommunicate between on-brane and off-brane equations.
We solve these equations iteratively until we get the converged perturbations.

The master variable $\Omega$ is parametrized by the exponent $p$ 
on the brane as
\ba
\Omega\mid_{y=0} = A(p) a^p\,.
\ea
To take this as an initial condition, we can set the separable function $G$
evolving through the bulk coordinate $y$ satisfying $G\mid_{y=0}=1$.
It is convenient to define a new variable $x\equiv yH$,
then the 5-dimensional scaling ansatz can be separable as,
\be
\Omega(a,x) = A(p) a^p G(x)\,.
\ee
The causal horizon of the propagation of perturbations through bulk is given by
\ba
\xi=aH^2\int^a_0\frac{da'}{a'^2H(a')^2}\,.
\ea
The second boundary condition can be imposed at $x=\xi$ as $G(x=\xi)=0$,
until it hits the horizon in Gaussian normal coordinate $y=1/H$ $(x=1)$.
We switch the boundary condition at $x=1$ at $\xi>1$.
It is fair to set this boundary condition since there are no other sources
to propagate fluctuations beyond those horizons.

We put this scaling ansatz into the off-brane equation Eq.~\ref{eq:master},
\ba\label{e:ODE}
A(x)\dbd{^2G}{x^2}+B(x)\dbd{G}{x}+C(x)G=0\,,
\ea
where the coefficients are given by
\ba
A(x)&=&(1-x)\left[1-x(1+2h)\right]\nn\\
B(x)&=&-2xhp-2xh^2-xh'-(1+h)+x(1+h)^2\nn\\
&-&x^2h\frac{h'+h^2+h}{1-x(h+1)}+3xh\frac{[1-x(1+h)]}{1-x}\nn\\
&+&3\frac{[1-x(1+h)]^2}{1-x}\nn\\
C(x)&=&-p^2-hp-xp\frac{h'+h^2+h}{1-x(h+1)}\nn\\
&+&3p\frac{1-x-xh}{1-x}-\frac{\left[1-x(1+h)\right]^2}{(1-x)^2}
\frac{k^2}{a^2H^2}\,,
\ea
where $h$ denotes $({\rm d}H/{\rm d\ln a})/H$.
When we solve Eq.~\ref{e:ODE}, it gives the gradient coefficient $R$
in terms of $G$,
\be
R=\frac{\frac{dG}{dx}\mid_{x=0}}{G\mid_{x=0}}\,.
\ee
Then we can solve the on-brane equation Eq.~\ref{eq:Bian}.

We start the iteration from solving the off-brane equation with 
the trial value $p$.
We use $p=4$ which is the expected value of $p$ during matter domination epoch.
It returns the value of $R$, and now the on-brane equation can be solved.
The on-brane equation gives more realistic value of $p$,
\ba
p=\frac{{\rm d\, ln}\Omega}{{\rm d\, ln }a}\,.
\ea
Then the same process is repeated until the solution converges.

\begin{figure}[t]%[htbp]
  \begin{center}
  \epsfysize=3.0truein
  \epsfxsize=3.0truein
    \epsffile{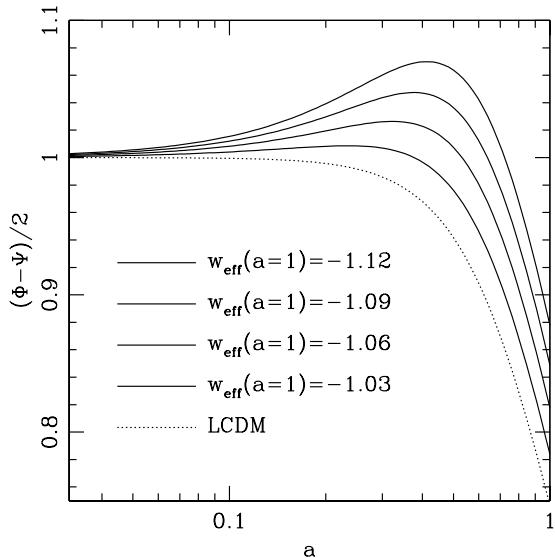}
    \caption{\footnotesize 
The perturbed potentials $(\Phi-\Psi)/2$ in quasi-static limit.
The effective equation of state $w_{\rm eff}$ varies with $\Omega_{\sigma}$
with fixing all other cosmological parameters.
The dotted curve represents $(\Phi-\Psi)/2$ of LCDM, and
the solid curves represent $(\Phi-\Psi)/2$ with varying $w_{\rm eff}$
from $-1.03$ to $-1.12$.
}
\label{fig:Quasi}
\end{center}
\end{figure}

In the quasi-static limit, the spatial gradient term 
dominates over all other quasi-statically varying terms.
Then $\pi_E$ is simply related to $\delta_E$ by constraint equation
to close on-brane equations of motion,
\ba
\pi_E=\frac{3}{2k^2}\delta_E\,.
\ea
The perturbed potentials can be written in the simple form,
\ba
\frac{k^2}{a^2}\Phi=4\pi G\left(1-\frac{1}{3\beta}\right)\rho\Delta_m
\ea
\ba
\frac{k^2}{a^2}\Psi=-4\pi G\left(1+\frac{1}{3\beta}\right)\rho\Delta_m \,,
\ea
where
\ba\label{eq:beta}
\beta=1+2Hr_c\left(1+\frac{\dot{H}}{3H^2}\right)\,.
\ea
The reversed sign in $\beta$ leads to $(1+1/3\beta)>1$.
Then the Newtonian potential well becomes deeper in nDGP than in LCDM,
and matter fluctuations falling into that well are enhanced.
It is the opposite behavior of perturbations
observed in sDGP where matter fluctuations decay
due to the shallower Newtonian potential well.
The dynamic solution from iteration routine precisely
reproduces this result.
The quasi-static limit is valid for nDGP.
We plot the solution in Fig.~\ref{fig:Quasi}
with varying $w_{\rm eff}$ from $-1.03$ to $-1.12$.

\begin{figure}[t]%[htbp]
  \begin{center}
  \epsfysize=3.0truein
  \epsfxsize=3.0truein
    \epsffile{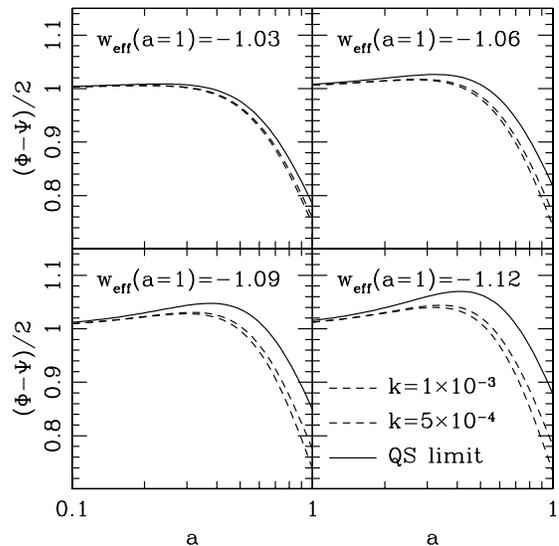}
    \caption{\footnotesize 
The perturbed potentials $(\Phi-\Psi)/2$ at scales of nearly horizon size
are plotted.
The solid curves represent the quasi static solutions,
and the dash curves represent the dynamic scaling solutions
scales at $k= 10^{-3} {\rm Mpc^{-1}}$ and $k=5\times 10^{-4} {\rm Mpc^{-1}}$.
}
\label{fig:scaling}
\end{center}
\end{figure}

For the modes in crossing the horizon, the quasi-static approximation
is invalid. Weyl fluid anisotropy is influenced by all other terms
including the bulk gradient.
The dynamic solution starts to depart from the solution of
the quasi-static approximation.
Its behavior is presented in Fig.~\ref{fig:scaling} at larger scales
of $k=10^{-3} {\rm Mpc}^{-1}$ and $k=5\times10^{-4} {\rm Mpc}^{-1}$. 
The dynamic solution departs from the quasi-static solution at scales of $k$
smaller than $k<10^{-2} {\rm Mpc}^{-1}$. 
The ISW effect on CMB power spectra is influenced by this departure,
while the ISW effect on ISW-galaxy cross-correlations is well described
by using quasi-static approximation only.

We test the numerical stability by using Bertschinger's equation.
Bertschinger derived the dynamic equation for the perturbed potential
at super-horizon modes based upon just two conditions, the Friedman universe
and the conservation of energy.
As long as these two conditions are not broken in the theory,
the following equation is always valid at the limit of 
$k\rightarrow 0$~\cite{Wands00,Bertschinger06},
\ba\label{eq:bert}
\Phi''-\Psi'-\frac{H''}{H'}\Phi'
+\left(\frac{H''}{H'}-\frac{H'}{H}\right)\Phi=0
\ea
here $'$ denotes the derivative in terms of $\ln a$.
We take the numerical anisotropy stress in the limit of $k\rightarrow 0$
and feed it into Eq.~\ref{eq:bert}.
The numerical perturbed potentials from the scaling method
are consistent with the solution from Eq.~\ref{eq:bert}.

\section{The validity of scaling method in asymptotic limit}
We derive the stability condition for nDGP in the asymptotic limit.
In the future, nDGP approaches to the 
de Sitter universe. All modes are expelled
outside the horizon, $\xi$ becomes infinity, and 
$k/(aH)^2 \ll 1$ limit is available for all modes.
The critical distance $r_c$ is greater than
the constant asymptotic particle horizon $1/H(\infty)$,
which makes $c\equiv H(\infty)r_c>1$.
In the de Sitter limit, the master equation is written as
\ba\label{eq:2}
\frac{d^2\Omega}{d y^2}
+\frac{2H}{1-Hy}\frac{d\Omega}{d y}
-\frac{H^2}{(1-Hy)^2}\left(\frac{d^2\Omega}{d \ln a^2}
-3\frac{d\Omega}{d \ln a}\right)=0\,.
\ea
Two boundary conditions, $\Omega(y=0)=1$ and $\Omega(y=1/H)=0$,
leads to the solutions trivially in the following forms,
\ba\label{eq:s}
\Omega=A(p)a^p\left[B_+(1-Hy)^{q_+}+B_-(1-Hy)^{q_-} \right]\,.
\ea 
Since $(1-Hy)$ is zero at $y=1/H$, the negative exponent of $(1-Hy)$ will
violate the normalizability. Thus only the positive exponent will be 
acceptable with $B_+=1$ to satisfy the other boundary condition at $y=0$.

The solution of master equation, Eq.~\ref{eq:s}, 
gives us the bulk gradient at $y=0$, $R=-q_+$.
Then the conservation equation of master variable on the brane 
in de Sitter limit is simply written as,
\ba
q_+^2+\frac{1}{c}q_+-\frac{1}{c}=0\,,
\ea
where we ignores the contribution of matter perturbations which
converges to constant at late times.
The positive solution $q_+$ is
\ba
q_+=-\frac{1}{2c}+\frac{1}{2}\sqrt{\frac{1}{c^2}+\frac{4}{c}}\,.
\ea
It gives the evolution equation of $\Omega$ on the brane at late times,
\ba
p^2-3p-\left[\frac{1}{c}-\left(2+\frac{1}{c}\right)q_+\right]=0\,.
\ea
The growing mode has the exponent $p$ as,
\ba\label{eq:p}
p=\frac{3}{2}+\half\sqrt{9+\frac{4}{c}-4q_+\left(2+\frac{1}{c}\right)}
\ea
The growing solution of $\Omega$ with this $p$
leads to the growing 
anisotropy. With the choice of $c\sim 2$, $\Phi_+$ becomes
\ba
\Phi_+\propto \frac{\pi_E}{a}\sim A_+a^{1.7}\,.
\ea

But it is a misleading conclusion of asymptotic behavior of nDGP
by using scaling method improperly at late times.
The negative $R$ violates the renormalizability condition~\cite{Koyama06mh}.
It indicates that the real exponent of $q$ can not be a solution,
and it should be imaginary.
Then we are not able to ignore $B_-$ by regularity condition at $y=1/H$.
Since the boundary condition at $y=1/H$ does not determine any coefficient
of $B$'s, the solution of nDGP at asymptotic region is unbounded.
Thus the scaling solution is invalidated in the de Sitter limit~\cite{Karoy}.
We can use the scaling method when the matter perturbations keep growing
and attract $\Omega$ to grow, up to the present time or near future until
hitting de Sitter limit.

\section{conclusion}
We have solved the dynamic equations of nDGP by using the scaling method.
A detectable departure from LCDM is found with $w_{\rm eff}$ below $-1$.
The matter fluctuations are enhanced due to the deeper potential well
in nDGP than in LCDM,
which weaken the ISW effect.
We are able to probe nDGP by detecting this weakened ISW effect
by ISW-galaxy or ISW-weak lensing cross-correlation.
It is highly recommended to select high redshift bins where
the modified perturbed potential increases to provide 
the negative cross-correlation.
If found, this would be strong evidence for the presence of
the nDGP model~\cite{Giannantonio07}.

\begin{acknowledgments}
We would like to thank Kelly Laas for reading this manuscript,
thank Kazuya Koyama for helpful comments on a draft of this
paper, and thank Antonio Cardoso, 
Nemanja Kaloper, Roy Maartens, Giuseppe De Risi, Ignacy Sawicki, 
Sanjeev Seahra and Fabio Silba for useful conversations.
This work is supported by STFC.
\end{acknowledgments}

%\bibliography{stableDGP}

\end{document}